\documentclass[12pt, preprint]{aastex}
\usepackage{natbib}
\shorttitle{Dark Matter Detection in Hard X-ray}

\shortauthors{Jeltema and Profumo}
\def\epem{$e^{\pm}$ }

\begin{document}

\title{Dark Matter Detection with Hard X-ray Telescopes}

\author{Tesla E. Jeltema\altaffilmark{1} and Stefano Profumo\altaffilmark{1}}

\altaffiltext{1}{Department of Physics and Santa Cruz Institute for Particle Physics,  University of California, 1156 High St., Santa Cruz, CA 95064, USA}

\begin{abstract}
We analyze the impact of future hard X-ray observations on the search for indirect signatures of particle dark matter in large extragalactic systems such as nearby clusters or groups of galaxies. We argue that the hard X-ray energy band falls squarely at the peak of the inverse Compton emission from electrons and positrons produced by dark matter annihilation or decay for a large class of dark matter models. Specifically, the most promising are low-mass models with a hard electron-positron annihilation final state spectrum and intermediate-mass models with a soft electron-positron spectrum.  We find that constraints on dark matter models similar to the current constraints from the Fermi Gamma-Ray Space Telescope will be close to the sensitivity limit of the near-term hard X-ray telescopes NuSTAR and ASTRO-H for relatively long observations.  An instrument like the Wide Field Imager (WFI) proposed for ATHENA would instead give a significant gain in sensitivity to dark matter if placed in a low background orbit similar to NuSTAR's; however, given the higher expected background level for ATHENA's proposed orbit at L2, its sensitivity will be similar to that of NuSTAR.  
\end{abstract}


\section{INTRODUCTION}

Well into its third year of science operations, the Fermi gamma-ray space telescope mission \citep{fermiatwood} has reached a high degree of maturity, and delivered very significant science results, including for the field of new physics and dark matter searches. Of the utmost importance for the latter was the detailed, survey-mode exploration of the gamma-ray sky in the 1-100 GeV energy range. In the near future, the high-energy sky will be probed with unprecedented accuracy by a new generation of hard X-ray space telescopes, including NuSTAR \citep{nustar}, ASTRO-H \citep{astroh} and Advanced Telescope for High ENergy Astrophysics (ATHENA) \citep{ixo}. In the present study, we investigate the role that future hard X-ray observations can play in the quest for the particle nature of dark matter.

The search for dark matter with the Fermi telescope has been an extremely active field within and outside the Fermi Collaboration. Unfortunately, no clear association of a gamma-ray signal with dark matter has been possible so far, thus only limits on properties of the dark matter particle could be derived. An incomplete list of the most significant results includes searches for an annihilation signature from nearby dwarf spheroidal galaxies \citep{dwarfstack,dsph}, for an annihilation or decay signal from nearby groups and clusters of galaxies \citep{fermidm, decay}, the search for monochromatic gamma-ray lines \citep{lines, Vertongen:2011mu} and the search for cosmological dark matter annihilation in the isotropic diffuse gamma-ray sky \citep[e.g.][]{diffuse, Papucci:2009gd}.  Atmospheric Cherenkov Telescopes have also undertaken an aggressive and comprehensive campaign of indirect searches for gamma rays from dark matter at higher energies than Fermi-LAT. For example, the H.E.S.S. experiment targeted the Galactic center \citep{Abramowski:2011hc}, the Sculptor, Carina \citep{:2010zzt} and Sagitarrius dwarf galaxies \citep{Aharonian:2007km}, while the MAGIC telescope recently reported searches for dark matter annihilation in the Perseus cluster of galaxies \citep{Aleksic:2009ir} and in the Segue I dwarf galaxy \citep{Aleksic:2011jx}.

Since the Fermi telescope is by design an electron-positron (\epem) detector, accurate spectral information on cosmic-ray \epem was also obtained \citep{epem1, epem2}. This measurement attracted wide-spread attention given the excess high-energy positron flux reported by the Pamela telescope \citep{pamela}, a possible hint of Galactic dark matter annihilation or decay, though Fermi gamma-ray observations also put strong constraints on these models \citep{fermidm, decay}. Claims of a gamma-ray signal from the region of the Galactic center that might be associated to dark matter annihilation have also been put forward \citep{hoopergc}. 
Clearly, while Fermi data allowed significant constraints to be put on dark matter models, several tantalizing observations possibly related to particle dark matter signals keep open the possibility that the high-energy sky might unveil particle physics beyond the Standard Model.

Clusters of galaxies have been prime targets for searches for dark matter annihilation and decay, for several reasons. First, they are the largest collapsed dark matter structures in the universe. Additionally, the sheer size of clusters of galaxies is such that secondary stable particles produced in dark matter annihilation, and in particular \epem, loose energy on time scales much shorter than the timescale for these particles to diffuse out of the system \citep[as is instead the case for smaller objects such as dwarf galaxies, see e.g.][]{dwarfus, CPU2}. This fact implies that a significant secondary emission from \epem resulting from dark matter annihilation and decay should be detectable from clusters, if the annihilation rate is large enough \citep{CPU1}. Indeed, some of the most stringent constraints on particle dark matter models that would explain the Pamela positron fraction with dark matter annihilation or decay have been set utilizing the predicted inverse Compton (IC) emission from the up-scattering of cosmic microwave background (CMB) photons by \epem from dark matter annihilation \citep{fermidm} or decay \citep{decay} in clusters.

Gamma rays are a prime tool to search for dark matter in the class of weakly interacting massive particles (WIMP) primarily because the gamma-ray energy range corresponds to the energy scale of the expected WIMP mass. However, future gamma-ray instruments will, presumably, focus on higher energies than Fermi \citep[for example with the advent of the Cherenkov Telescope Array (CTA),][]{cta}: this will be beneficial to test TeV-scale dark matter models, but will leave substantially out of reach low-mass dark matter candidates, i.e. WIMPs with masses on the order of a few GeV. As we briefly review in what follows, several tantalizing experimental data, primarily from direct dark matter detection experiments, point towards this class of WIMP models. However, the existence of a multi-wavelength dark matter emission spectrum, that includes the (primary) prompt emission of radiation from the annihilation or decay event as well as the secondary emission from stable charged particles also produced by dark matter annihilation or decay, can be exploited to constrain dark matter models.

We point out here that for low-mass WIMP models (roughly, WIMPs lighter than half a TeV) the peak of the IC emission expected for example from clusters of galaxies falls squarely in the energy range that will be probed by the next generation hard X-ray telescopes. Such relatively light WIMPs are not only of interest in view of recent experimental developments, but also because theoretical arguments indicate that particle dark matter models that are being -- and will be -- tested with the Large Hadron Collider (LHC) should have masses precisely in that range. Hard X-ray experiments could therefore play an important role in the forthcoming years, possibly complementary to direct dark matter detection and to collider searches for new physics.

In the present study we outline in sec.~\ref{sec:theory} theoretical arguments that indicate that IC emission from light WIMPs peaks in the hard X-ray band making this a very interesting energy range to search for a dark matter signal, especially in environments like galaxy clusters, but possibly also in the Galactic center. We illustrate this with a few selected WIMP models. As an example, in sec.~\ref{sec:calc} we calculate the sensitivity of NuSTAR, slated to launch in February 2012, to the relevant hard X-ray signals and derive the performance of NuSTAR over the WIMP parameter space of pair-annihilation rate versus mass, discussing also the anticipated performance of other proposed X-ray telescopes like ASTRO-H and ATHENA. 
Finally, we present our discussion and conclusions in sec.~\ref{sec:concl}.

\section{DARK MATTER SIGNATURES IN HARD X-RAY}\label{sec:theory}

We argue in this Section that the hard X-ray band is a very promising energy range to search for annihilation or decay of weakly interacting massive particles. In particular, we show that the energy range probed by upcoming telescopes like NuSTAR, ASTRO-H, and ATHENA covers the location of the Inverse Compton (IC) peak expected from the up-scattering of background photons by electrons and positrons (\epem) produced by weak-scale particle dark matter annihilation or decay.

The average  frequency $\langle \nu\rangle$ of IC up-scattered photons with an initial frequency $\nu_0$ by an electron or positron (\epem) with a Lorentz factor $\gamma_{\rm el}$ is \citep{longair}:
\begin{equation}
\langle \nu\rangle\approx \left(\frac{4}{3}\gamma_{\rm el}^2\right)\ \nu_0.
\end{equation}
Relevant frequencies for us are those of cosmic microwave background photons, $\nu_0\sim 10^{-3}$ eV, of starlight photons, $\nu_0\sim 1$ eV, and of IR photons resulting from the scattering, absorption and reemission of the starlight by dust, $\nu_0\sim 10^{-2}$ eV. The typical Lorentz factor of \epem resulting from the annihilation or decay of dark matter depend on the dark matter mass $m_\chi$ and on the annihilation final state. Approximately, for typical hard ($\mu^+\mu^-$) and soft ($b\bar b$) annihilation or decay final states, one finds
\begin{eqnarray}
\nonumber\gamma_{\rm el}\sim\left(\frac{m_\chi}{2}\right)\frac{1}{m_e} & {\rm for\ a} & \mu^+\mu^-\ {\rm final\ state};\\
\nonumber\gamma_{\rm el}\sim\left(\frac{m_\chi}{20}\right)\frac{1}{m_e} & {\rm for\ a} & b\bar b\ {\rm final\ state}.
\end{eqnarray}
As a result, a rule of thumb for the energy of  photons up-scattered by \epem produced by a dark matter particle with a mass $m_\chi$ gives
\begin{equation}
\langle \nu\rangle\approx\left(\frac{m_\chi}{10}\frac{1}{m_e}\right)^2\nu_0.
\end{equation}
Thus, we find that typically
\begin{eqnarray}
&&\langle \nu\rangle\sim 1-100\ \left(\frac{m_\chi}{10\ \rm GeV}\right)^2\  {\rm keV}\quad {\rm (CMB)}\label{eq:cmb}\\
&&\langle \nu\rangle\sim 10-1000\ \left(\frac{m_\chi}{10\ \rm GeV}\right)^2\  {\rm keV}\quad {\rm (Dust)}\\
&&\langle \nu\rangle\sim 1-100\ \left(\frac{m_\chi}{10\ \rm GeV}\right)^2\  {\rm MeV}\quad {\rm (Starlight)}
\end{eqnarray}
In systems like clusters or groups of galaxies, where dark matter annihilation or decay occurs across the entire dark matter halo, CMB photons dominate the overall target photon energy, thus the bulk of the IC emission is expected precisely across the energy band relevant for NuSTAR, ASTRO-H and for the Wide Field Imager (WFI) proposed for ATHENA.

\begin{figure}
\begin{center}
\includegraphics[width=0.7\textwidth]{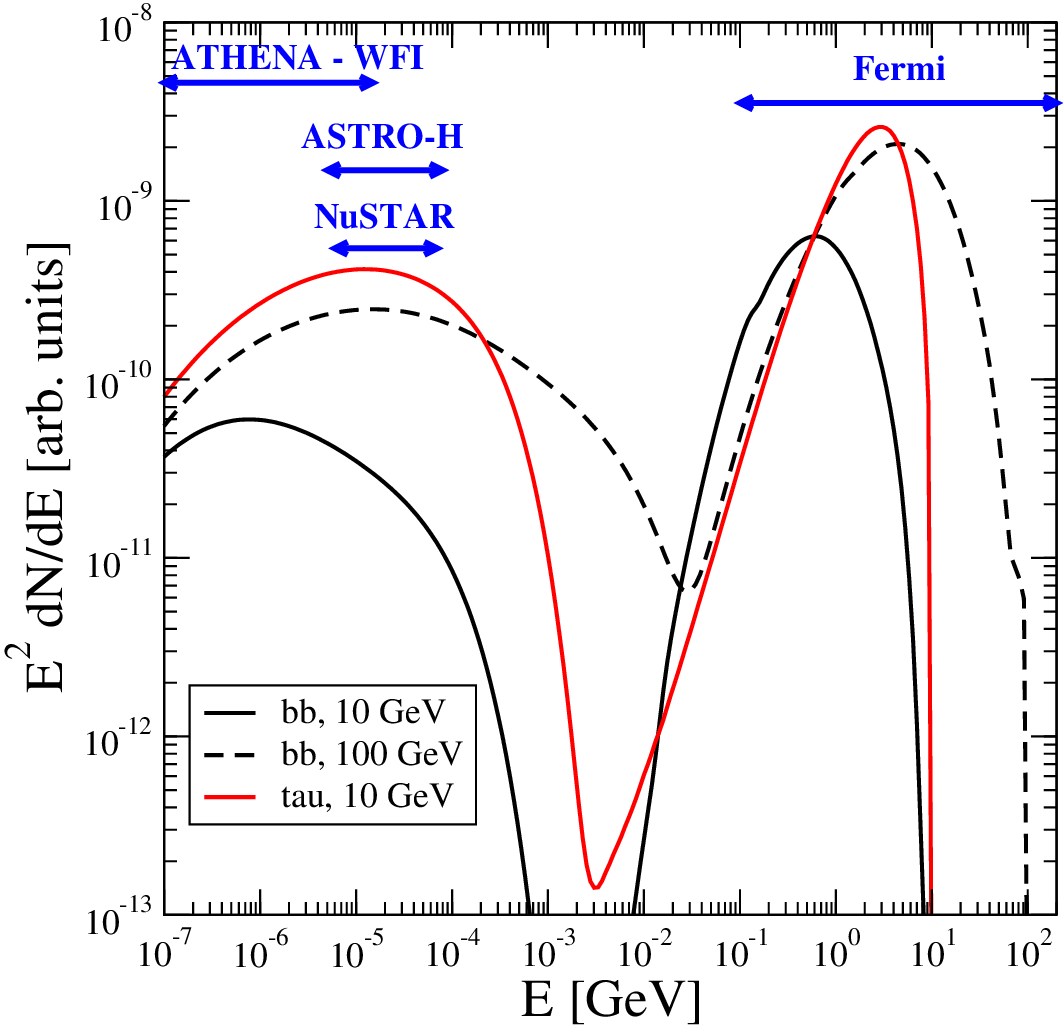}
\end{center}
\caption{Shown, in arbitrary units, but normalized to the same integrated flux above 0.1 GeV, is the spectral energy distribution of dark matter models with a $b\bar b$ annihilation final state and a mass of 10 (solid black line) and 100 GeV (black dashed line), and with a $\tau^+\tau^-$ final state and a mass of 10 GeV (red solid line). We also indicate, with horizontal blue arrows, the relevant energy range for NuSTAR, ASTRO-H, WFI on ATHENA, and the Fermi gamma-ray space telescope.}
\label{fig:bbbar}
\end{figure}

We illustrate in Fig.~\ref{fig:bbbar} the multi-wavelength spectrum expected from the annihilation of weakly-interacting dark matter particles (WIMPs) \citep[for related early studies see e.g.][]{CPU1, CPU2, PU}. In the figure we show, with arbitrary units, the spectral energy distribution (SED) for three generic, model-independent WIMP spectra. Horizontal lines indicate the approximate energy range relevant for NuSTAR, for the Fermi Gamma-Ray Space Telescope, for ASTRO-H, and for the WFI on ATHENA. For the sake of illustration, we choose to normalize the spectra to have the same integrated photon flux above 0.1 GeV. Also, we conservatively only include CMB photons as target background radiation for IC scattering: higher energy photons from starlight and dust would produce a further, albeit subdominant feature at higher energies than the IC peak shown in our spectra. In addition, we neglect diffusion of \epem based on the fact that the typical timescale for \epem diffusion at scales comparable to those of groups or clusters of galaxies is much longer than the energy loss timescale for the \epem energies of interest here \citep[see e.g. fig.~A.3 of ][]{CPU1}.

In Fig.~\ref{fig:bbbar} we choose two models whose dominant pair-annihilation mode is into a pair of $b\bar b$ quarks, with two different particle masses of 10 and 100 GeV, and one model with a $\tau^+\tau^-$ annihilation final state and a particle mass of 10 GeV. Notice that since dark matter is believed to be non-relativistic in the Galaxy, the SEDs are truncated at a photon energy equivalent to the dark matter particle mass by conservation of energy and momentum.

We now briefly comment on the physical motivation behind the choices of particle dark matter models. The $b\bar b$ final state is typical of certain classes of WIMP models, including supersymmetry \citep{susyrev}; the gamma-ray and \epem yield of the $b\bar b$ final state is, additionally, very similar to that of a generic quark-antiquark or gluon-gluon final state, and is, in this respect, a quite representative choice. 

The choice of a 10 GeV mass is representative of the mass range preferred by recent direct detection experimental claims, including CoGeNT \citep{cogent}, DAMA/LIBRA \citep{dama} and CRESST-II \citep{cresst}, though whether these experimental results might or might not be related to the direct detection of WIMPs remains highly debated and controversial. Also, the ``crossing symmetry'' problem of reconstructing the pair-annihilation final state from direct detection results is well beyond the current experimental information, so any guess constitutes the result of some theoretical prior.  The choice of a 100 GeV mass is, instead, motivated by the lightest possible neutralino mass in minimal supersymmetric extensions of the Standard Model (MSSM) with gaugino mass unification at the grand unification scale, in view of recent null searches for supersymmetry with the Large Hadron Collider \citep[see e.g.][and also see results presented at the 2011 summer conferences by the ATLAS and CMS Collaborations]{Aad:2011hh, Khachatryan:2011tk}. Again, the final state choice depends on the bias towards supersymmetric dark matter models with a neutralino as the lightest supersymmetric particle.

Finally, we chose a 10 GeV candidate with a $\tau^+\tau^-$ final state in view of recent claims that observations including gamma rays from the center of the Galaxy \citep{hoopergc}, the radio excess known as the WMAP haze \citep{hooperhaze} and the synchrotron emission from Galactic Center radio filaments \citep{hooperfilaments} might be related to the pair-annihilation of a 7-10 GeV WIMP dominantly annihilating into leptons, and specifically into $\tau^+\tau^-$ pairs. While these claims are also rather controversial, a $\tau^+\tau^-$ annihilation final state is plausible in some WIMP models, including for example the coannihilation region of the constrained MSSM. Examples of dark matter models with a preferential $\tau^+\tau^-$ annihilation final states have also been recently been considered by \cite{taitetal, heather, boucenna}.

Notice that the gamma-ray and the \epem production in the case of $b\bar b$ (or, generically, of strongly interacting) annihilation final states stems primarily from the decays of neutral and charged pions produced in the hadronization of the quark decay products. These secondary pions have significantly lower energies than the primary decaying particle, resulting in a soft gamma-ray and \epem spectrum. In the case of $\tau^+\tau^-$ final state a similar production mechanism exists for the $\tau$ hadronic decay channels, but much harder photons and \epem are produced by the leptonic decay modes into muons and electrons, as well as from internal bremsstrahlung. The  $\tau^+\tau^-$ final state thus produces a {\em harder} gamma-ray and IC spectrum, for a given particle dark matter mass.

While the particular choice of low masses (10 GeV) we employ for two of the three illustrative spectra is motivated by possibly controversial recent data, we argue that the class of annihilation final states we consider are indeed quite representative of theoretically motivated WIMP models, and have thus great generality. Indeed, in supersymmetry, barring fine-tuned regions of parameter space where the scalar-tau is almost degenerate with the lightest neutralino (stau coannihilation region), the gamma-ray and \epem spectra correspond to either quark-antiquark final states, or to gauge boson pairs (in the case of higgsino- or wino-like neutralinos). The (dominant) hadronic decay channels for massive gauge bosons  produce spectra for \epem very similar to what we consider here -- the electrons and positrons originate from high-energy charged pion decay chains in both cases. In this respect, the spectral features of the $b\bar b$ final state can be regarded, from a theoretical perspective as highly representative. We also argue that the $\tau^+\tau^-$ is theoretically well-motivated. First, it can be viewed as the only exception to the $b\bar b$ final state class discussed above, for the stau coannihilation region, in the context of supersymmetric dark matter models; secondly, it reproduces rather closely what is expected in another theoretically motivated class of dark matter models: universal extra dimensions \citep{Hooper:2007qk}. For this reason, in what follows we concentrate on the two choices of $b\bar b$ and $\tau^+\tau^-$ annihilation final states for the spectra of \epem and gamma rays from dark matter annihilation.

Fig.~\ref{fig:bbbar} shows that the detailed spectral predictions closely match the approximations outlined in the early part of this section, in particular Eq.~(\ref{eq:cmb}), with rather representative hard and soft \epem spectra corresponding to the $\tau^+\tau^-$ and to the $b\bar b $ final states, respectively. Also, the scaling with the WIMP mass given again in Eq.~(\ref{eq:cmb}) indicates that for supersymmetric WIMP candidates, which preferentially pair-annihilate into final states leading to soft \epem spectra, the hard X-ray regime will cover the peak of the IC emission for virtually all neutralino models with GUT-scale gaugino mass unification that will be tested with the LHC.  Depending on the number of years the LHC will collect data for, it is reasonable to expect a sensitivity to gluinos as heavy as 2-3 TeV; gaugino mass unification implies that a bino-like lightest neutralino have a mass $\sim1/6$ the mass of the gluino, implying WIMP masses up to 300-500 GeV. What we find in Fig.~\ref{fig:bbbar} implies that even at 500 GeV a significant IC emission would fall within the NuSTAR energy range.

We conclude this section by remarking that all of the considerations above hold for the case of a meta-stable dark matter scenario where the dark matter particle decays with a macroscopic lifetime. In this case, however, the resulting spectra for the case of the pair annihilation of a particle of mass $m_{\rm ann}$ annihilating into a Standard Model final state $XX$ corresponds to the spectrum of a particle of mass $m_{\rm dec}=2\times m_{\rm ann}$ decaying into the same final state $XX$, by conservation of energy.

\section{DARK MATTER DETECTION WITH PROPOSED HARD X-RAY TELESCOPES}\label{sec:calc}

In this section, we specifically investigate the potential for NuSTAR observations of nearby clusters of galaxies to reveal or constrain a signal from dark matter annihilation.  The effective area of the Hard X-ray Telescope (HXT) on ASTRO-H\footnote{http://astro-h.isas.jaxa.jp/researchers/sim/effective\_area.html} will be very comparable to that of NuSTAR and its sensitivity to dark matter similar.  The WFI proposed for ATHENA
, while primarily a soft X-ray instrument, would give a large gain in effective area at energies less than $\sim10-12$ keV compared to NuSTAR and current X-ray telescopes like XMM and Chandra, though with higher background than the former, and we also estimate its potential sensitivity to a dark matter signal.  

We simulated the NuSTAR sensitivity for a 1 Msec observation of a nearby cluster using the background and response files provided by the NuSTAR team\footnote{http://www.NuSTAR.caltech.edu/for-astronomers/simulations}.  Specifically, we estimate the flux limit which can be obtained by NuSTAR for an extended source in three energy bands, 6-10 keV, 10-30 keV, and 30-60 keV.  We consider a source region with a radius of six arcminutes, and the NuSTAR background and response files were modified appropriately for a source of this size.  At off-axis angles greater than 6', the NuSTAR effective area is less than half of the on-axis effective area\footnote{See the NuSTAR Performance Guide at http://www.NuSTAR.caltech.edu/for-astronomers/simulations}; in addition the predicted dark matter annihilation signal is proportional to the square of the dark matter density and drops quickly with cluster radius. The combination of these effects means that little signal is expected for NuSTAR at larger radii.

We model the hard X-ray spectrum from dark matter annihilation induced IC emission as an exponentially cut-off power law with a photon index of 1.5 and a cut-off energy of around 100 keV. The expected spectrum, in particular the energy cut-off, depends on the specific dark matter model and particle mass considered, but we found that the NuSTAR photon flux limits depend only very weakly on changes in the spectrum for the range of dark matter models we consider.  We find that for a 1 Msec observation, NuSTAR could detect a cluster flux of roughly $1.1 \times 10^{-6}$ photons cm$^2$ s$^{-1}$ in the 6-10 keV band at three sigma above the background; the flux limits are $2.5 \times 10^{-6}$ photons cm$^2$ s$^{-1}$ and $5.4 \times 10^{-6}$ photons cm$^2$ s$^{-1}$ for the 10-30 keV and 30-60 keV bands, respectively.

In the above calculations, we do not consider any additional background which might come from astrophysical cluster X-ray emission.  Below we will consider the specific case of a hard X-ray observation of the Fornax cluster, which was found to be the best candidate cluster for detecting a gamma-ray signal from dark matter annihilation or decay with Fermi \citep{clustersus, fermidm, decay}.  Fornax is a low-mass cluster/group of galaxies with a low temperature thermal X-ray gas  \citep[][]{scharf, buote}.  Conservatively taking the temperature of the hotter thermal component found by \cite{buote} with $kT=1.5$ keV, we find that the thermal gas in the Fornax cluster would contribute a count rate of $\sim$1\% of the X-ray background for NuSTAR in the 6-10 keV band and negligibly to harder bands.  We add this component as an additional background to dark matter detection in what follows. Due to their proximity, nearby groups of galaxies (such as e.g. Fornax, M49, and NGC4636) have given the strongest individual constraints on dark matter models based on the non-detection of gamma-ray emission with Fermi \citep{fermidm, decay}, and these would be good targets for hard X-ray searches as well.  For larger, hotter clusters, thermal X-ray emission could contribute a significant flux at the lowest energies probed by NuSTAR. 

For some nearby clusters additional sources of background could come from non-thermal hard X-ray emission (e.g. debated non-thermal emission in the Coma cluster \citep{wik}) or bright, central AGN (e.g. NGC1275 in Perseus and M87 in Virgo), but again these sources are not present in some of the best candidate groups and clusters like Fornax.  Hard X-ray emission has only been detected from a handful of clusters which are typically merging clusters with detected large-scale diffuse radio emission.  The origin of the hard X-ray emission is sometimes ascribed to IC emission by the cosmic ray electrons generating the radio halo, but in most cases can simply arise from shock heated, multi-temperature gas \citep[e.g.][]{ajello1,ajello2}.  Many clusters, however, do not show evidence for non-thermal emission in that they lack detected diffuse radio emission \citep[e.g.][]{brunetti}, limiting the possible cosmic ray electron population and consequently the possible non-thermal hard X-ray emission. Fornax does not have detected radio halo emission, shows no evidence for gas hotter than $\sim$ 2 keV, and has not been detected in the Swift-BAT multi-year all-sky survey \citep{scharf, ajello2}. NuSTAR is an imaging hard X-ray telescope, and its spatial resolution will be sufficient to remove any background AGN detected.

\begin{figure}
\begin{center}
\includegraphics[width=0.7\textwidth]{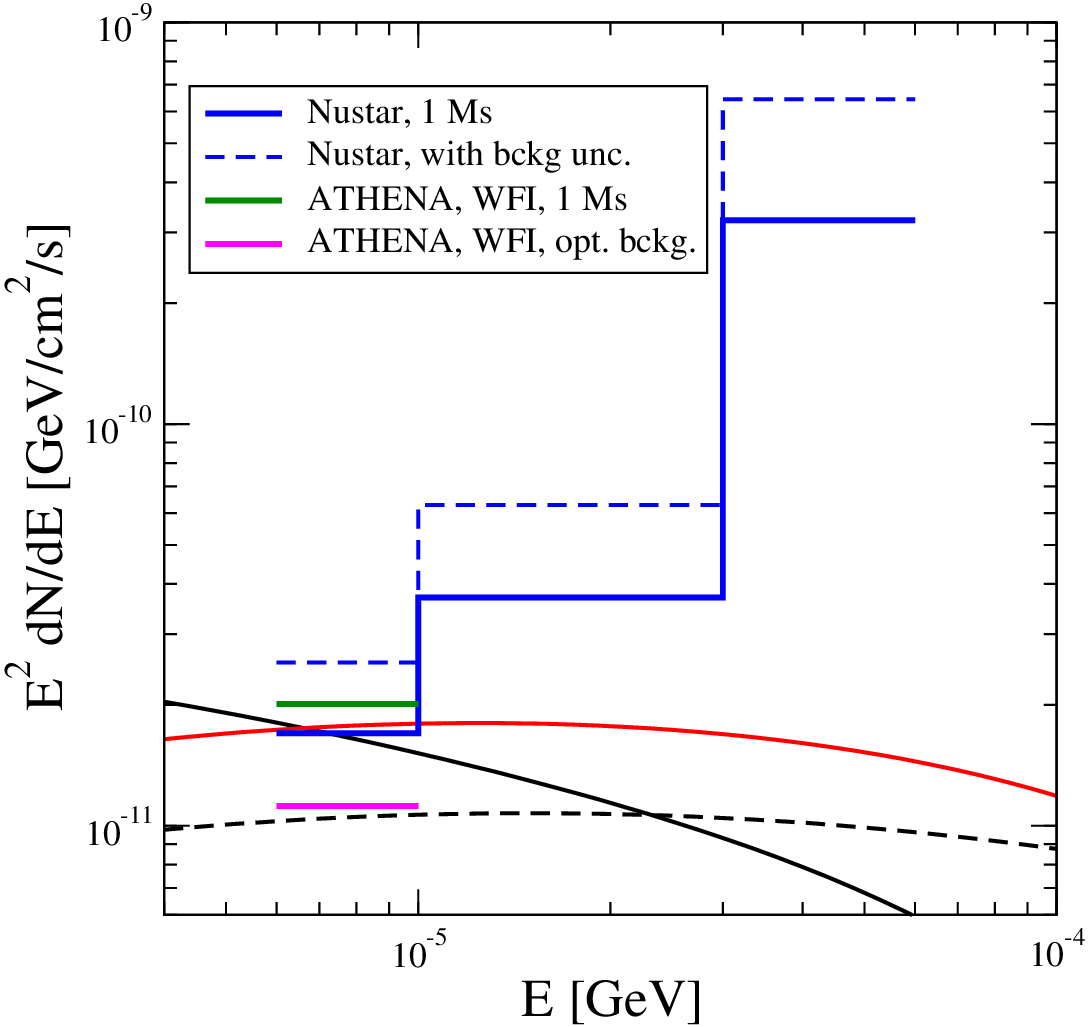}
\end{center}
\caption{Shown is the hard X-ray spectral energy distribution of the same models as in fig.~\ref{fig:bbbar}, here normalized to the current Fermi limits for the Fornax cluster. We also show the NuSTAR sensitivity in three energy bins for a 1 Msec observations (solid blue line).  The dashed blue line shows the sensitivity including possible uncertainty in the NuSTAR background rate.  The green line shows the predicted sensitivity of the WFI instrument proposed for ATHENA in the 6-10 keV band also for a 1 Msec observation.  The magenta line instead shows the sensitivity that an instrument like the WFI would have if placed in a similar orbit to NuSTAR where the NXB is considerably lower than at L2, the orbit chosen for ATHENA.}
\label{fig:sed_nustar}
\end{figure}

Fig.~\ref{fig:sed_nustar} illustrates the potential sensitivity of NuSTAR to a dark matter annihilation signal. Specifically, the predicted emission for the same dark matter models as in the previous section is shown in an extended X-ray band, this time normalized to the limits from the Fermi Gamma-ray Space Telescope on the same dark matter models from observations of the Fornax cluster of galaxies \citep{fermidm}. Note that the field-of-view of NuSTAR is small compared to that of the Fermi-LAT. NuSTAR, with the assumed source size of 6' radius, would probe the central, dark matter densest cluster regions and would detect a local cluster as an extended source.  Fermi, on the other hand, has a point spread function of around one degree at the most relevant energies for dark matter detection, making even nearby clusters essentially point-like.  In Fig.~\ref{fig:sed_nustar}, we have renormalized the curves to reflect the difference in the expected dark matter annihilation flux from the inner 6' of the Fornax cluster compared to the flux from a radius of $1^{\circ}$, which is roughly a factor 2.3 larger for an assumed NFW or Einasto profile. The plot thus allows a direct comparison between the performance of Fermi as a dark matter detector with that of near term hard X-ray telescopes. We compare the predictions with the expected sensitivity for NuSTAR, shown with blue lines 
and calculated as described above.  The solid line shows the expected sensitivity for a 1 Msec observation.  The X-ray background level for NuSTAR has been estimated by the NuSTAR team based on simulations and observations\footnote{See the NuSTAR Performance Guide at http://www.NuSTAR.caltech.edu/for-astronomers/simulations} \citep{harrison}, but the actual background level may be different.  In Fig.~\ref{fig:sed_nustar}, the dashed blue line shows an estimate of the impact of systematic uncertainty in the background level on the potential dark matter constraints.  Here we include a factor of 2 uncertainty in the non-X-ray (particle) background level\footnote{http://www.inta.es/g4suw2009/docs/Presentations/20\_Wed\_May\_2009/G4SUWS\_2009\_Zoglauer\_NuSTARBackgroundAndG4Activation.pdf} and a 20\% uncertainty in the cosmic X-ray background level \citep{gilli}.  An additional source of uncertainty will come from the precision to which the actual X-ray background can be determined once NuSTAR is operational.  Luckily, compared to the soft X-ray background, the hard X-ray background is spatially uniform, and with current telescopes like Chandra the background above a few keV is predictable to the couple percent level\footnote{http://cxc.harvard.edu/contrib/maxim/acisbg/COOKBOOK}.  However, this level of systematic uncertainty is similar to the dark matter flux one would like to measure (in the case of the solid blue limits shown in Figure 2, the dark matter count rate is a little more than 1\% of the total background in the 6-10 keV band), and this level of uncertainty in the background measurement would weaken the dark matter constraints.

\begin{figure}
\begin{center}
\includegraphics[width=0.7\textwidth]{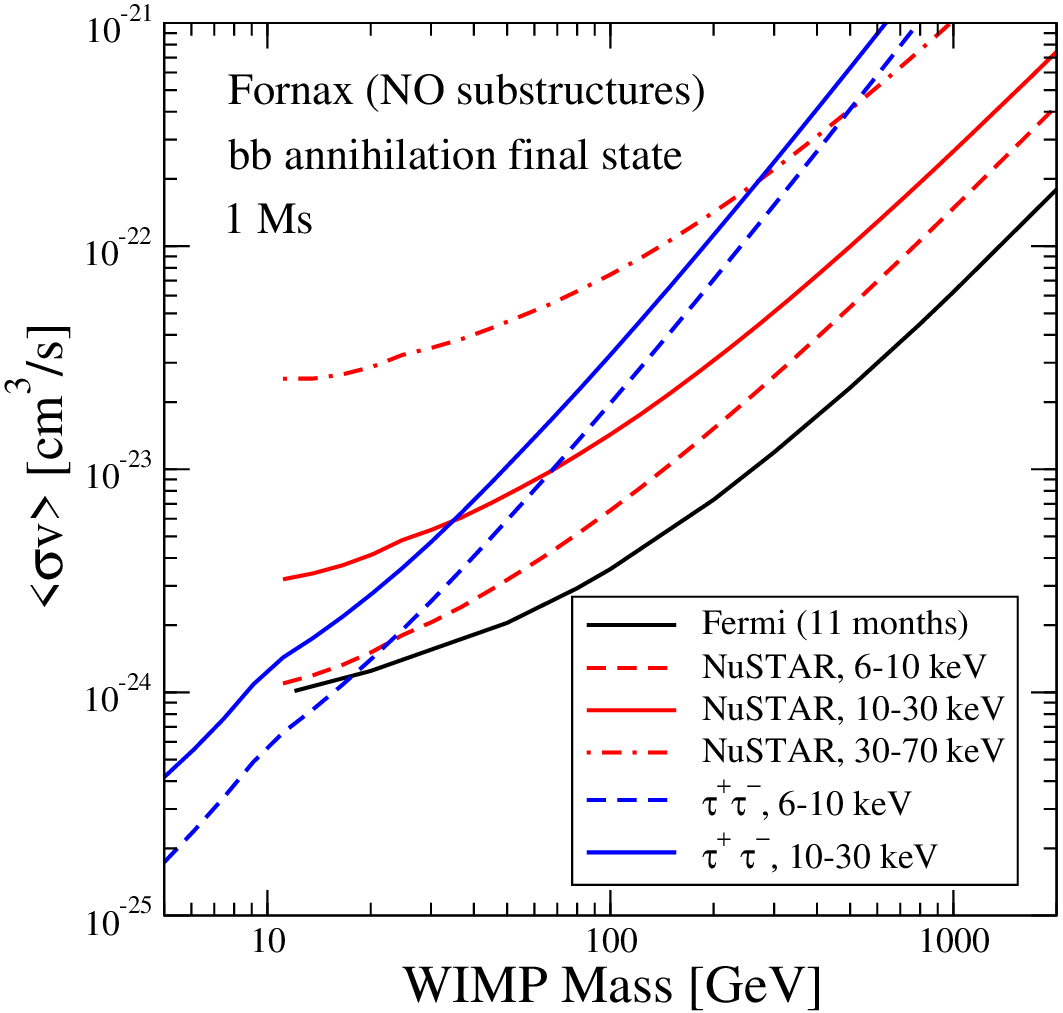}
\end{center}
\caption{Shown is a comparison of the performance of NuSTAR (in three distinct energy bands, red lines) and of the Fermi gamma-ray space telescope (solid black line) for the detection of a dark matter signal from annihilation in the Fornax cluster, for a $b\bar b$ final state, on the pair-annihilation versus mass parameter space. We also show the NuSTAR sensitivity to a $\tau^+\tau^-$ annihilation final state, for the 6-10 keV and 10-30 keV energy bands.}
\label{fig:mx}
\end{figure}

We translate in Fig.~\ref{fig:mx} our predictions for the sensitivity of NuSTAR as a dark matter detector onto the plane of the WIMP pair-annihilation cross section versus mass, again for the Fornax cluster. The black line represents the published Fermi limits for a $b\bar b$ final state \citep[see][]{fermidm}.  All lines show the constraints without accounting for any substructure in the cluster. Notice that this is clearly an overly conservative assumption, as galaxies within the cluster enhance the annihilation signal by a factor of a few compared to the signal from the smooth halo alone \citep{fermidm, clustersus}. However, rather than the absolute value of the pair annihilation cross section that hard X-ray constraints will allow us to probe, here we are interested in comparing the performance of future X-ray telescopes with current gamma-ray observations.

The predictions for the sensitivity of a 1 Msec NuSTAR observation of the core of Fornax to WIMP annihilation to a $b\bar b$ final state are shown with red lines: the dashed line corresponds to the softest, 6-10 keV energy band, for which we include background contamination from the cluster thermal bremsstrahlung emission (very small for Fornax, but potentially important for other clusters); the solid line corresponds to the intermediate 10-30 keV energy band, and the dot-dashed line to the highest energy band, between 30 and 60 keV.  We also indicate the sensitivity to a hard pair-annihilation final sate (here, again, $\tau^+\tau^-$) with blue lines. The convention is again that the dashed line corresponds to the 6-10 keV energy band, and the solid line to the intermediate 10-30 keV band. Given the interest in low-mass WIMP models with a $\tau^+\tau^-$ mentioned in the previous section, we show here also masses below 10 GeV.

Fig.~\ref{fig:sed_nustar} and Fig.~\ref{fig:mx} show that a long NuSTAR observation of the Fornax cluster in the 6-10 keV energy range would yield similar constraints to those from Fermi for very low dark matter particle masses around 10 GeV for both a $b\bar b$ and a $\tau^+\tau^-$ final state, if the background level is similar to the nominal level predicted.  For larger masses, the NuSTAR constraints would be about a factor of two weaker for the $b\bar b$ case.  The constraints weaken somewhat if the NuSTAR background is systematically higher.

In the most constraining energy band (6-10 keV), an instrument like the WFI proposed for ATHENA would give a large gain in effective area.  
However, the current plan for ATHENA is to place it at L2 where the non-X-ray (particle) background is predicted to be significantly higher than for the low Earth orbit selected for NuSTAR \citep{ixobkgd}.  Using the currently available response and background files\footnote{http://sci.esa.int/science-e/www/object/index.cfm?fobjectid=43974}, we find that combining these factors the ATHENA WFI would have a similar sensitivity to dark matter annihilation in Fornax as that of NuSTAR, which we show as the green line in Fig.~\ref{fig:sed_nustar}.  If, however, the non-X-ray background for ATHENA was similar to that of NuSTAR (i.e. similar orbit), its sensitivity would be significantly better and stronger than the Fermi limits for low dark matter particle masses (magenta line in Fig.~\ref{fig:sed_nustar}).  While this estimate does not correspond to a planned mission, it does show that with near term technology, one could design an X-ray mission with the ability to place interesting constraints on dark matter models.  In these estimates, we include the expected thermal cluster emission.

We stress that a result very similar to what is presented in Fig.~\ref{fig:mx} would correspond to the case of meta-stable dark matter decay, with the pair annihilation cross section replaced by the (inverse) lifetime, and with a rescaling of a factor 2 of the WIMP mass on the x-axis. A possible difference would correspond to the amount of matter within the line of sight, the relevant quantity for decay signals, which might, depending upon the dark matter density distribution, favor gamma-ray observations with a larger field of view over hard X-ray observations, though a mosaic of X-ray observations could be used to cover a larger cluster area.

\section{DISCUSSION AND CONCLUSIONS}\label{sec:concl}

We considered the performance of future hard X-ray telescopes in searching for indirect signals from low-mass dark matter particles, a class of models that will be out of reach for future gamma ray telescopes such as CTA. We carried out detailed estimates for the NuSTAR telescope sensitivity to the annihilation signal in the Fornax cluster of galaxies from three well-motivated benchmark dark matter models. We concluded that only with a relatively long observing time will NuSTAR get close to the current sensitivity of the Fermi gamma-ray telescope to the same models. We also pointed out that the Wide Field Instrument proposed for ATHENA will achieve a similar sensitivity level to NuSTAR in for its current proposed orbit, but that an instrument like the WFI placed in a low background orbit (like that chosen for NuSTAR) could exceed the Fermi sensitivity for low mass WIMP models.  While novel constraints on dark matter may be a bit out of reach of upcoming hard X-ray telescopes, the fact that for a large range of dark matter models the expected secondary IC emission form dark matter annihilation peaks in this band means that dark matter searches can be very relevant to the planning of future X-ray missions.  In particular, an appropriately planned mission with a low background orbit, a high sensitivity around 10 keV, and a large field-of-view could provide unique information on light weakly-interacting massive particle dark matter models.

In this study we did not focus on the imaging capabilities of future hard X-ray instruments, which will be significantly better than what is possible with present and future gamma-ray telescopes. While in the observation of distant extragalactic objects the instrumental angular resolution is much less critical than the instrument field of view in the search for a signal from dark matter, there are other astrophysical environments where the opposite might be true. A clear example is the Galactic center, where source confusion plagues the possibility of a solid discrimination between a diffuse signal from dark matter versus multiple point sources with gamma-ray observations. Hard X-ray data on the Galactic center region will greatly help clarify the nature of the high-energy emission from that region, in particular in connection with a possible dark matter interpretation.


\acknowledgments
This work is partly supported by NASA grants NNX09AT96G and NNX09AT83G. SP acknowledges support from an Outstanding Junior Investigator Award from the Department of Energy, DE-FG02-04ER41286.

\bibliography{nustar}

\begin{thebibliography}{49}
\expandafter\ifx\csname natexlab\endcsname\relax\def\natexlab#1{#1}\fi

\bibitem[{Aad {et~al.}(2011)}]{Aad:2011hh}
Aad, G., et al. 2011, Phys.Rev.Lett., 106, 131802

\bibitem[{Aalseth {et~al.}(2011)}]{cogent}
Aalseth, C., et al. 2011, Phys.Rev.Lett., 106, 131301

\bibitem[{Abdo {et~al.}(2010{\natexlab{a}})Abdo, Ackermann, Ajello, Atwood,
  Baldini, {et~al.}}]{lines}
Abdo, A., Ackermann, M., Ajello, M., Atwood, W., Baldini, L., et al.
  2010{\natexlab{a}}, Phys.Rev.Lett., 104, 091302

\bibitem[{Abdo {et~al.}(2010{\natexlab{b}})Abdo, Ackermann, Ajello, Atwood,
  Baldini, {et~al.}}]{dsph}
---. 2010{\natexlab{b}}, Astrophys.J., 712, 147

\bibitem[{Abdo {et~al.}(2010{\natexlab{c}})}]{diffuse}
Abdo, A., et al. 2010{\natexlab{c}}, JCAP, 1004, 014

\bibitem[{Abdo {et~al.}(2009)}]{epem1}
Abdo, A.~A., et al. 2009, Phys.Rev.Lett., 102, 181101

\bibitem[{Abramowski {et~al.}(2011{\natexlab{a}})}]{:2010zzt}
Abramowski, A., et al. 2011{\natexlab{a}}, Astropart.Phys., 34, 608

\bibitem[{Abramowski {et~al.}(2011{\natexlab{b}})}]{Abramowski:2011hc}
---. 2011{\natexlab{b}}, Phys.Rev.Lett., 106, 161301

\bibitem[{{Ackermann} {et~al.}(2011){Ackermann}, {Ajello}, {Albert}, {Atwood},
  {Baldini}, {Ballet}, {Barbiellini}, {Bastieri}, {Bechtol}, {Bellazzini},
  {Berenji}, {Blandford}, {Bloom}, {Bonamente}, {Borgland}, {Bregeon},
  {Brigida}, {Bruel}, {Buehler}, {Burnett}, {Buson}, {Caliandro}, {Cameron},
  {Canadas}, {Caraveo}, {Casandjian}, {Cecchi}, {Charles}, {Chekhtman},
  {Chiang}, {Ciprini}, {Claus}, {Cohen-Tanugi}, {Conrad}, {Cutini}, {de
  Angelis}, {de Palma}, {Dermer}, {Digel}, {Silva}, {Drell}, {Drlica-Wagner},
  {Falletti}, {Favuzzi}, {Fegan}, {Ferrara}, {Fukazawa}, {Funk}, {Fusco},
  {Gargano}, {Gasparrini}, {Gehrels}, {Germani}, {Giglietto}, {Giordano},
  {Giroletti}, {Glanzman}, {Godfrey}, {Grenier}, {Guiriec}, {Gustafsson},
  {Hadasch}, {Hayashida}, {Hays}, {Hughes}, {Jeltema}, {Johannesson},
  {Johnson}, {Johnson}, {Kamae}, {Katagiri}, {Kataoka}, {Kn{\"o}dlseder},
  {Kuss}, {Lande}, {Latronico}, {Lionetto}, {Llena Garde}, {Longo}, {Loparco},
  {Lott}, {Lovellette}, {Lubrano}, {Madejski}, {Mazziotta}, {McEnery},
  {Mehault}, {Michelson}, {Mitthumsiri}, {Mizuno}, {Monte}, {Monzani},
  {Morselli}, {Moskalenko}, {Murgia}, {Naumann-Godo}, {Norris}, {Nuss},
  {Ohsugi}, {Okumura}, {Omodei}, {Orlando}, {Ormes}, {Ozaki}, {Paneque},
  {Parent}, {Pesce-Rollins}, {Pierbattista}, {Piron}, {Pivato}, {Porter},
  {Profumo}, {Raino}, {Razzano}, {Reimer}, {Reimer}, {Ritz}, {Roth},
  {Sadrozinski}, {Sbarra}, {Scargle}, {Schalk}, {Sgro}, {Siskind}, {Spandre},
  {Spinelli}, {Strigari}, {Suson}, {Tajima}, {Takahashi}, {Tanaka}, {Thayer},
  {Thayer}, {Thompson}, {Tibaldo}, {Tinivella}, {Torres}, {Troja}, {Uchiyama},
  {Vandenbroucke}, {Vasileiou}, {Vianello}, {Vitale}, {Waite}, {Wang}, {Winer},
  {Wood}, {Wood}, {Yang}, {Zimmer}, {Kaplinghat}, \& {Martinez}}]{dwarfstack}
{Ackermann}, M., {et~al.} 2011, ArXiv e-prints

\bibitem[{Ackermann {et~al.}(2010{\natexlab{a}})Ackermann, Ajello, Allafort,
  Baldini, Ballet, {et~al.}}]{fermidm}
Ackermann, M., Ajello, M., Allafort, A., Baldini, L., Ballet, J., et al.
  2010{\natexlab{a}}, JCAP, 1005, 025, * Temporary ent ry *

\bibitem[{Ackermann {et~al.}(2010{\natexlab{b}})}]{epem2}
Ackermann, M., et al. 2010{\natexlab{b}}, Phys.Rev., D82, 092004

\bibitem[{Adriani {et~al.}(2009)}]{pamela}
Adriani, O., et al. 2009, Nature, 458, 607

\bibitem[{Aharonian(2008)}]{Aharonian:2007km}
Aharonian, .~F. 2008, Astropart.Phys., 29, 55, * Brief entry *

\bibitem[{{Ajello} {et~al.}(2009){Ajello}, {Rebusco}, {Cappelluti}, {Reimer},
  {B{\"o}hringer}, {Greiner}, {Gehrels}, {Tueller}, \& {Moretti}}]{ajello1}
{Ajello}, M., {et~al.} 2009, \apj, 690, 367

\bibitem[{{Ajello} {et~al.}(2010){Ajello}, {Rebusco}, {Cappelluti}, {Reimer},
  {B{\"o}hringer}, {La Parola}, \& {Cusumano}}]{ajello2}
{Ajello}, M., {Rebusco}, P., {Cappelluti}, N., {Reimer}, O., {B{\"o}hringer},
  H., {La Parola}, V., \& {Cusumano}, G. 2010, \apj, 725, 1688

\bibitem[{Aleksic {et~al.}(2010)}]{Aleksic:2009ir}
Aleksic, J., et al. 2010, Astrophys.J., 710, 634

\bibitem[{Aleksic {et~al.}(2011)}]{Aleksic:2011jx}
---. 2011, JCAP, 1106, 035

\bibitem[{Angloher {et~al.}(2011)Angloher, Bauer, Bavykina, Bento, Bucci,
  {et~al.}}]{cresst}
Angloher, G., Bauer, M., Bavykina, I., Bento, A., Bucci, C., et al. 2011

\bibitem[{Atwood {et~al.}(2009)}]{fermiatwood}
Atwood, W., et al. 2009, Astrophys.J., 697, 1071

\bibitem[{Barcons {et~al.}(2011)Barcons, Barret, Bautz, Bookbinder, Bregman,
  {et~al.}}]{ixo}
Barcons, X., Barret, D., Bautz, M., Bookbinder, J., Bregman, J., et al. 2011

\bibitem[{Bernabei {et~al.}(2010)Bernabei, Belli, Cappella, Cerulli, Dai,
  {et~al.}}]{dama}
Bernabei, R., Belli, P., Cappella, F., Cerulli, R., Dai, C., et al. 2010,
  Eur.Phys.J., C67, 39

\bibitem[{Boucenna \& Profumo(2011)}]{boucenna}
Boucenna, M., \& Profumo, S. 2011

\bibitem[{{Brunetti} {et~al.}(2009){Brunetti}, {Cassano}, {Dolag}, \&
  {Setti}}]{brunetti}
{Brunetti}, G., {Cassano}, R., {Dolag}, K., \& {Setti}, G. 2009, \aap, 507, 661

\bibitem[{Buckley {et~al.}(2010)Buckley, Hooper, \& Tait}]{taitetal}
Buckley, M.~R., Hooper, D., \& Tait, T.~M. 2010

\bibitem[{{Buote}(2002)}]{buote}
{Buote}, D.~A. 2002, \apjl, 574, L135

\bibitem[{Colafrancesco {et~al.}(2006)Colafrancesco, Profumo, \& Ullio}]{CPU1}
Colafrancesco, S., Profumo, S., \& Ullio, P. 2006, Astron.Astrophys., 455, 21

\bibitem[{Colafrancesco {et~al.}(2007)Colafrancesco, Profumo, \& Ullio}]{CPU2}
---. 2007, Phys.Rev., D75, 023513

\bibitem[{Dugger {et~al.}(2010)Dugger, Jeltema, \& Profumo}]{decay}
Dugger, L., Jeltema, T.~E., \& Profumo, S. 2010, JCAP, 1012, 015

\bibitem[{{Gilli} {et~al.}(2007){Gilli}, {Comastri}, \& {Hasinger}}]{gilli}
{Gilli}, R., {Comastri}, A., \& {Hasinger}, G. 2007, \aap, 463, 79

\bibitem[{{Harrison} {et~al.}(2010){Harrison}, {Boggs}, {Christensen}, {Craig},
  {Hailey}, {Stern}, {Zhang}, {Angelini}, {An}, {Bhalereo}, {Brejnholt},
  {Cominsky}, {Cook}, {Doll}, {Giommi}, {Grefenstette}, {Hornstrup}, {Kaspi},
  {Kim}, {Kitaguchi}, {Koglin}, {Liebe}, {Madejski}, {Kruse Madsen}, {Mao},
  {Meier}, {Miyasaka}, {Mori}, {Perri}, {Pivovaroff}, {Puccetti}, {Rana}, \&
  {Zoglauer}}]{harrison}
{Harrison}, F.~A., {et~al.} 2010, in Society of Photo-Optical Instrumentation
  Engineers (SPIE) Conference Series, Vol. 7732, Society of Photo-Optical
  Instrumentation Engineers (SPIE) Conference Series

\bibitem[{Harrison {et~al.}(2010)Harrison, Boggs, Christensen, Craig, Hailey,
  {et~al.}}]{nustar}
Harrison, F.~A., Boggs, S., Christensen, F., Craig, W., Hailey, C., et al. 2010

\bibitem[{Hermann(2011)}]{cta}
Hermann, G. 2011, Nucl.Phys.Proc.Suppl., 212-213, 170

\bibitem[{Hooper \& Goodenough(2011)}]{hoopergc}
Hooper, D., \& Goodenough, L. 2011, Phys.Lett., B697, 412

\bibitem[{Hooper \& Linden(2011)}]{hooperhaze}
Hooper, D., \& Linden, T. 2011, Phys.Rev., D83, 083517

\bibitem[{Hooper \& Profumo(2007)}]{Hooper:2007qk}
Hooper, D., \& Profumo, S. 2007, Phys.Rept., 453, 29

\bibitem[{Jeltema {et~al.}(2009)Jeltema, Kehayias, \& Profumo}]{clustersus}
Jeltema, T.~E., Kehayias, J., \& Profumo, S. 2009, Phys.Rev., D80, 023005

\bibitem[{{Jeltema} \& {Profumo}(2008)}]{dwarfus}
{Jeltema}, T.~E., \& {Profumo}, S. 2008, \apj, 686, 1045

\bibitem[{Jungman {et~al.}(1996)Jungman, Kamionkowski, \& Griest}]{susyrev}
Jungman, G., Kamionkowski, M., \& Griest, K. 1996, Phys.Rept., 267, 195

\bibitem[{Khachatryan {et~al.}(2011)}]{Khachatryan:2011tk}
Khachatryan, V., et al. 2011, Phys.Lett., B698, 196

\bibitem[{Linden {et~al.}(2011)Linden, Hooper, \&
  Yusef-Zadeh}]{hooperfilaments}
Linden, T., Hooper, D., \& Yusef-Zadeh, F. 2011

\bibitem[{Logan(2011)}]{heather}
Logan, H.~E. 2011, Phys.Rev., D83, 035022

\bibitem[{{Longair}(2010)}]{longair}
{Longair}, M.~S. 2010, {High Energy Astrophysics}, ed. {Longair, M.~S.}

\bibitem[{Papucci \& Strumia(2010)}]{Papucci:2009gd}
Papucci, M., \& Strumia, A. 2010, JCAP, 1003, 014

\bibitem[{Profumo \& Ullio(2010)}]{PU}
Profumo, S., \& Ullio, P. 2010

\bibitem[{{Scharf} {et~al.}(2005){Scharf}, {Zurek}, \& {Bureau}}]{scharf}
{Scharf}, C.~A., {Zurek}, D.~R., \& {Bureau}, M. 2005, \apj, 633, 154

\bibitem[{{Smith} {et~al.}(2010){Smith}, {Bautz}, {Bookbinder}, {Garcia},
  {Guainazzi}, \& {Kilbourne}}]{ixobkgd}
{Smith}, R.~K., {Bautz}, M.~W., {Bookbinder}, J., {Garcia}, M.~R., {Guainazzi},
  M., \& {Kilbourne}, C.~A. 2010, in Society of Photo-Optical Instrumentation
  Engineers (SPIE) Conference Series, Vol. 7732, Society of Photo-Optical
  Instrumentation Engineers (SPIE) Conference Series

\bibitem[{Takahashi {et~al.}(2010)Takahashi, Mitsuda, Kelley, Aharonian,
  Akimoto, {et~al.}}]{astroh}
Takahashi, T., Mitsuda, K., Kelley, R., Aharonian, F., Akimoto, F., et al.
  2010, Proc.SPIE Int.Soc.Opt.Eng., 7732, 77320Z

\bibitem[{Vertongen \& Weniger(2011)}]{Vertongen:2011mu}
Vertongen, G., \& Weniger, C. 2011, JCAP, 1105, 027

\bibitem[{{Wik} {et~al.}(2011){Wik}, {Sarazin}, {Finoguenov}, {Baumgartner},
  {Mushotzky}, {Okajima}, {Tueller}, \& {Clarke}}]{wik}
{Wik}, D.~R., {Sarazin}, C.~L., {Finoguenov}, A., {Baumgartner}, W.~H.,
  {Mushotzky}, R.~F., {Okajima}, T., {Tueller}, J., \& {Clarke}, T.~E. 2011,
  \apj, 727, 119

\end{thebibliography}

\end{document}